\documentclass[12pt]{article}
\usepackage{geometry} 

\usepackage{graphicx}
\usepackage{tabularx}
\usepackage{typearea}
\usepackage{wrapfig}
\usepackage{fancyhdr}
\usepackage{amsmath}
\usepackage{amssymb}
\usepackage{amsthm}
\usepackage[ansinew]{inputenc}
\usepackage[arrow, matrix, curve]{xy}
\usepackage{enumerate}
\usepackage{bbm}
\usepackage{mathtools}
\usepackage{cite}
\usepackage{tikz}
\usetikzlibrary{calc}

\theoremstyle{theorem}
\newtheorem{Theorem}{Theorem}[section]
\newtheorem{Proposition}[Theorem]{Proposition}
\newtheorem{Lemma}[Theorem]{Lemma}
\newtheorem{Corollary}[Theorem]{Corollary}
\newtheorem{Definition}[Theorem]{Definition}

\theoremstyle{remark}
\newtheorem{Remark}[Theorem]{Remark}
\newtheorem{Example}[Theorem]{Example}

\newcommand{\btm}{\begin{Theorem}}
\newcommand{\etm}{\end{Theorem}}

\newcommand{\ben}{\begin{enumerate}}
\newcommand{\een}{\end{enumerate}}

\newcommand{\bit}{\begin{itemize}}
\newcommand{\eit}{\end{itemize}}

\newcommand{\bca}{\begin{cases}}
\newcommand{\eca}{\end{cases}}

\newcommand{\bre}{\begin{Remark}\rm}
\newcommand{\ere}{\end{Remark}}

\newcommand*{\bbm}{\begin{Remark}}
\newcommand*{\ebm}{\end{Remark}}

\newcommand{\ble}{\begin{Lemma}}
\newcommand{\ele}{\end{Lemma}}

\newcommand*{\bsz}{\begin{Proposition}}
\newcommand*{\esz}{\end{Proposition}}

\newcommand{\beq}{\begin{equation}}
\newcommand{\eeq}{\end{equation}}

\newcommand{\bbma}{\begin{bmatrix}}
\newcommand{\ebma}{\end{bmatrix}}

\newcommand*{\bbs}{\begin{Example}}
\newcommand*{\ebs}{\end{Example}}

\newcommand*{\bfg}{\begin{Corollary}}
\newcommand*{\efg}{\end{Corollary}}

\newcommand*{\bdf}{\begin{Definition}}
\newcommand*{\edf}{\end{Definition}}

\newcommand*{\bbw}{\begin{proof}}
\newcommand*{\ebw}{\end{proof}}

\newcommand*{\bpf}{\begin{proof}}
\newcommand*{\epf}{\end{proof}}

\newtagform{simple}{}{}
\newtagform{standard}{(}{)}
\newcommand{\eqqed}{\usetagform{simple}\tag{$\square$}}
\newcommand{\eqqedan}{\usetagform{simple}}
\newcommand{\eqqedaus}{\usetagform{standard}}

\allowdisplaybreaks

\newcommand{\II}{\mathbbm{1}}

\newcommand{\CC}{{\mathbb{C}}}

\newcommand{\NN}{{\mathbb{N}}}

\newcommand{\N}{\mathbb{N}}

\newcommand{\bem}[1]{\text{#1}}
\newcommand{\momap}{\mu}

\newcommand*{\res}{\upharpoonright}

\newcommand{\sitem}{\rm\item\it}

\newcommand{\SL}{{\mr{SL}}}
\renewcommand{\sl}{\mf{sl}}

\newcommand{\SU}{{\mr{SU}}}
\newcommand{\su}{\mf{su}}

\newcommand{\al}[1]{\begin{align} #1 \end{align}}
\newcommand{\ala}[1]{\begin{align*} #1 \end{align*}}

\DeclareMathOperator{\Ad}{Ad}

\DeclareMathOperator{\im}{im}

\DeclareMathOperator{\tr}{tr}

\newcommand{\mc}[1]{\mathcal{#1}}
\newcommand{\mf}[1]{\mathfrak{#1}}
\newcommand{\mr}[1]{\mathrm{#1}}

\newcommand{\comment}[1]{}
\newcommand{\verweis}[1]{}
\newcommand{\todo}[1]{}

\newcommand{\pha}{{\mc P}}

\newcommand{\ve}{\varepsilon}

\newcommand{\ctg}{\mr T^\ast}

\newcommand{\rref}[1]{{\rm \ref{#1}}}

\newcommand{\ol}[1]{\overline{#1}}
\newcommand{\ul}[1]{\underline{#1}}
\newcommand{\abs}{\hspace*{2.5mm}}
\newcommand*{\qeb}{\nopagebreak\hspace*{0.1em}\hspace*{\fill}{\mbox{\small$\blacklozenge$}}}

\DeclareMathOperator{\ad}{ad}

\newcommand{\punkt}[1]{\put(#1){\circle*{0.06}}}

\newcommand{\vpunkte}[1]{\put(#1){\multiput(0,-0.1)(0,0.1){3}{\circle*{0.01}}}}

\newcommand{\linie}[3]{\put(#1){\line(#2){#3}}}

\newcommand{\whole}[3]{\put(#1){\punkt{0,0}\put(0.05,0.1){\makebox(-0.1,
   -0.2)[#2]{$#3$}}}}

\newcommand{\plene}[3]{\put(#1){\punkt{0,0}\linie{0.1,0}{1,0}{0.8}
\put(0.05,0.1){\makebox(-0.1,-0.2)[#2]{$#3$}}}}

\newcommand{\plvee}[3]{\put(#1){\punkt{0,0}\linie{0.0968,0.0242}{4,1}{0.8064}
\put(0.05,0.1){\makebox(-0.1,-0.2)[#2]{$#3$}}}}

\newcommand{\plvmee}[3]{\put(#1){\punkt{0,0}\linie{0.0968,-0.0242}{4,-1}{
0.8064}\put(0.05,0.1){\makebox(-0.1,-0.2)[#2]{$#3$}}}}

\newcommand{\pslene}[1]{\put(#1){\multiput(0.1,0)(0.1,0){9}{\circle*{0.01}}}}

\begin{document}

\title{Defining relations for the orbit type strata of $\SU(2)$-lattice gauge models}

 \author{
F.\ F\"urstenberg$^\dagger$, G.\ Rudolph$^\ast$, M.\ Schmidt$^\ast$
\\[5pt]
$^\dagger$ Physikalisches Institut, Universit\"at Freiburg
\\
Hermann-Herder-Str.\ 3, 79104 Freiburg, Germany
\\[5pt]
$^\ast$ Institut f\"ur Theoretische Physik, Universit\"at Leipzig
\\
Augustusplatz 10/11, 04109 Leipzig, Germany
 }

\date{\today}

\maketitle

\begin{abstract}

\noindent
We consider an $\SU(2)$-lattice gauge model in the tree gauge. Classically, this is a system with symmetries whose configuration space is a direct product of copies of $\SU(2)$, acted upon by diagonal inner automorphisms. We derive defining relations for the orbit type strata in the reduced classical phase space. The latter is realized as a certain quotient of a direct product of copies of the complexified group $\SL(2,\CC)$ (sometimes named the GIT-quotient because it provides a categorical quotient in the sense of geometric invariant theory). The relations derived can be used for the construction of the orbit type costratification of the Hilbert space of the quantum system in the sense of Huebsch\-mann. 

\end{abstract}

\section{Introduction}

This paper is part of a program which aims at developing a non-perturbative approach to the quantum theory of gauge fields in the Hamiltonian framework with special emphasis on the role of non-generic gauge orbit types. The starting point is a finite-dimensional Hamiltonian lattice approximation of the theory which leads, on the classical level, to a finite-dimensional Hamiltonian system with symmetries. On quantum level, we have investigated the observable algebra, constructed via canonical quantization and reduction, and its  superselection structure \cite{qcd1, qcd2, qcd3, RS}. For a first step towards the construction of the thermodynamical limit, see \cite{GR,GR2}. 

If the gauge group is non-Abelian, the action of the symmetry group in the corresponding Hamiltonian system with symmetries necessarily has more than one orbit type. Correspondingly, the reduced phase space, obtained by symplectic reduction, is a stratified symplectic space \cite{SjamaarLerman,OrtegaRatiu} rather than a symplectic manifold as in the case with one orbit type \cite{AbrahamMarsden}. The stratification is induced by the orbit types. It consists of an open and dense principal stratum and several secondary strata. Each of these strata is invariant under the dynamics with respect to any invariant Hamiltonian. See \cite{cfg, cfgtop,FRS} for case studies. 

Given that orbit type strata are a rather prominent feature on the classical level, the question arises whether they produce quantum effects. To investigate this, one can use the costratification of the Hilbert space of the quantum system in the sense of Huebsch\-mann \cite{Hue:Quantization} which is associated with the orbit type stratification of the reduced classical phase space. It is given by a family of closed subspaces, one for each stratum. Loosely speaking, the closed subspace associated with a certain stratum consists of the wave functions which are optimally localized at that stratum in the sense that they are orthogonal to all states vanishing at that stratum. The vanishing condition can be given sense in the framework of holomorphic quantization, where wave functions are true functions and not just classes of functions. In \cite{HRS} we have constructed this costratification for a toy model with gauge group $\SU(2)$ on a single lattice plaquette. As physical effects, we have found a nonzero tunneling probability between distant strata and, for a certain range of the coupling, a very large transition probability between the ground state of the lattice Hamiltonian and one of the two secondary strata.

The aim of the present paper is to make a step towards extending the results of
\cite{HRS} to $\SU(2)$-gauge models on arbitrary finite lattices. For that
purpose, we derive the defining relations for the orbit type strata in the
classical phase space. These are necessary for constructing the corresponding
closed subspaces. The explicit construction of these subspaces and their orthoprojectors remains as a future task.

\section{The model}

Let $G$ be a compact Lie group and let $\mf g$ be its Lie algebra. Later on, we will specify $G = \SU(2)$, but for the time being, this is not ncessary. Let $\Lambda$ be a finite spatial lattice and let $\Lambda^0$, $\Lambda^1$ and $\Lambda^2$ denote, respectively, the sets of lattice sites, lattice links and lattice plaquettes. For the links and plaquettes, let an arbitrary orientation be chosen. In lattice gauge theory with gauge group $G$ in the Hamiltonian approach, gauge fields (the variables) are approximated by their parallel transporters along links and gauge transformations (the symmetries) are approximated by their values at the lattice sites. Thus, the classical configuration space is the space $G^{\Lambda^1}$ of mappings $\Lambda^1 \to G$, the classical symmetry group is the group $G^{\Lambda^0}$ of mappings $\Lambda^0 \to G$ with pointwise multiplication and the action of $g \in G^{\Lambda^0}$ on $a \in G^{\Lambda^1}$ is given by  
\beq\label{G-Wir-voll}
(a \cdot g)(\lambda) := g(x) a(\lambda) g(y)^{-1}\,,
\eeq
where $\lambda \in \Lambda^1$ and $x$, $y$ denote the starting point and the endpoint of $\lambda$, respectively. The classical phase space is given by the associated Hamiltonian $G$-manifold \cite{AbrahamMarsden,Buch} and the reduced classical phase space is obtained from that by symplectic reduction \cite{OrtegaRatiu,Buch,SjamaarLerman}. We do not need the details here. Let us just mention that dynamics is ruled by the Kogut-Suskind lattice Hamiltonian. When identifying $\ctg G$ with $G \times \mf g$, and thus $\ctg G^{\Lambda^1}$ with $G^{\Lambda^1} \times \mf g^{\Lambda^1}$, by means of left-invariant vector fields, this Hamiltonian is given by 
$$
H(a,A)
 = 
\frac{g^2}{2 \delta} \sum_{\lambda \in \Lambda^1}^N \|A(\lambda)\|^2
 -
\frac{1}{g^2 \delta} \sum_{\pi \in \Lambda^2} \left(\tr a(\pi) + \ol{\tr a(\pi)}\right)
 \,,~
a \in G^{\Lambda^1},~ A \in \mf g^{\Lambda^1},
$$
where $g$ denotes the coupling constant, $\delta$ denotes the lattice spacing and $a(\pi)$ denotes the product of $a(\lambda)$ along the boundary of $\pi$ in the induced orientation. The trace is taken in some chosen unitary representation. Unitarity ensures that the Kogut-Suskind lattice Hamiltonian does not depend on the choice of plaquette orientations. 

When dicussing orbit types in continuum gauge theory, it is convenient to first factorize with respect to the free action of pointed gauge transformations, thus arriving at an action of the compact gauge group $G$ on the quotient manifold. This preliminary reduction can also be carried out in the case of lattice gauge theory under consideration. In fact, given a lattice site $x_0$, it is not hard to see that the normal subgroup \beq\label{G-ptgautrf}
\{g \in G^{\Lambda^0} : g(x_0) = \II\}\,,
\eeq
where $\II$ denotes the unit element of $G$, acts freely on $G^{\Lambda^1}$. Hence, one may pass to the quotient manifold and the residual action by the quotient Lie group of $G^{\Lambda^0}$ with respect to this normal subgroup. Clearly, the quotient Lie group is  naturally isomophic to $G$. The quotient manifold can be identified with a direct product of copies of $G$ and the quotient action can be identified with the action of $G$ by diagonal conjugation as follows. Choose a maximal tree $T$ in the graph $\Lambda^1$ and define the tree gauge of $T$ to be the subset
$$
\{a \in G^{\Lambda^1} : a(\lambda) = \II ~ \forall \, \lambda \in T\}
$$
of $G^{\Lambda^1}$. One can readily see that every element of $G^{\Lambda^1}$ is conjugate under $G^{\Lambda^0}$ to an element in the tree gauge of $T$ and that two elements in the tree gauge of $T$ are conjugate under $G^{\Lambda^0}$ if they are conjugate under the action of $G$ via constant gauge transformations. This implies that the natural inclusion mapping of the tree gauge into $G^{\Lambda^1}$ descends to a $G$-equivariant diffeomorphism from that tree gauge onto the quotient manifold of $G^{\Lambda^1}$ with respect to the action of the subgroup \eqref{G-ptgautrf}. Finally, by choosing a numbering of the off-tree links in $\Lambda^1$, we can identify the tree gauge of $T$ with the direct product of $N$ copies of $G$, where $N$ denotes the number of off-tree links. This number does not depend on the choice of $T$. Then, the action of $G$ on the tree gauge via constant gauge transformations translates into the action of $G$ on $G^N$ by diagonal conjugation,
\beq\label{G-Wir-Q}
g \cdot (a_1,\dots,a_N) = (g a_1 g^{-1} , \dots , g a_N g^{-1})\,.
\eeq
As a consequence of these considerations, for the discussion of the role of orbit types we may pass from the original large Hamiltonian system with symmetries, given by the configuration space $G^{\Lambda^1}$, the symmetry group $G^{\Lambda^0}$ and the action \eqref{G-Wir-voll}, to the smaller Hamiltonian system with symmetries given by the configuration space 
$$
Q := G^N\,,
$$
the symmetry group $G$ and the action of $G$ on $Q$ given by diagonal conjugation \eqref{G-Wir-Q}. This is the system we will discuss here. As before, the classical phase space is given by the associated Hamiltonian $G$-manifold and the reduced classical phase space is obtained from that by symplectic reduction. One can show that the latter is isomorphic, as a stratified symplectic space, to the reduced classical phase space defined by the original Hamiltonian system with symmetries. 

We will need the following information about the classical phase space. As a space, it is given by the cotangent bundle 
$$
\ctg Q \equiv \ctg G^N\,.
$$
It is a general fact that the action of $G$ on $Q$ naturally lifts to a symplectic action on $\ctg Q$ (consisting of the corresponding 'point transformations' in the language of canonical transformations) and that the lifted action admits a momentum mapping
$$
\mu : \ctg Q \to \mf g^\ast
 \,,~~~~~~
\mu(p)\big(A) := p(A_\ast)\,,
$$
where $p \in \ctg Q$, $A \in \mf g$ and $A_\ast$ denotes the Killing vector field defined by $A$. An easy calculation shows that under the global trivialization
\beq\label{G-trviz-N}
\ctg G^N \cong G^N \times \mf g^N
\eeq
induced by left-invariant vector fields and an invariant scalar product on $\mf g$, the lifted action is given by diagonal conjugation,
\beq\label{G-Wir-P}
g \cdot (a_1,\dots,a_N,A_1,\dots,A_N)
 = 
\big(g a_1 g^{-1} , \dots , g a_N g^{-1} , \Ad(g) A_1 , \dots , \Ad(g) A_N\big)
\eeq
and the associated momentum mapping is given by 
\beq\label{G-ImpAbb}
\mu(a_1,\dots,a_N,A_1,\dots,A_N)
 =
\sum_{i = 1}^N \Ad(a_i) A_i - A_i\,,
\eeq
see e.g.\ \cite[\S 10.7]{Buch}. The reduced phase space $\pha$ is obtained from $\ctg G^N$ by singular symplectic reduction at $\mu = 0$. That is, $\pha$ is the set of orbits of the lifted action of $G$ on the invariant subset $\momap^{-1}(0) \subseteq \ctg Q$, endowed with the quotient topology induced from the relative topology on this subset. In lattice gauge theory, the condition $\mu=0$ corresponds to the Gau{\ss} law constraint. As a matter of fact, the action of $G$ on $\momap^{-1}(0)$ has the same orbit types as that on $Q$. By definition, the orbit type strata of $\pha$ are the connected components of the subsets of $\pha$ of elements with a fixed orbit type. They are called strata because they provide a stratification of $\pha$ \cite{SjamaarLerman,OrtegaRatiu}. By the procedure of symplectic reduction, the orbit type strata of $\pha$ are endowed with symplectic manifold structures. The bundle projection $\ctg Q \to Q$ induces a mapping $\pha \to Q/G$. This mapping is surjective, because $\momap$ is linear on the fibres of $\ctg Q$ and hence $\momap^{-1}(0)$ contains the zero section of $\ctg Q$. It need not preserve the orbit type though.

\bbm

The tree gauge of $T$ need not be invariant under time evolution with respect to
a gauge-invariant Hamiltonian (e.g., the Kogut-Suskind lattice Hamiltonian), but
every motion in the full configuration space $G^{\Lambda^1}$ can be transformed
by a time-dependent gauge transformation to the tree gauge. Thus, up to
time-dependent gauge transformations, the tree gauge is invariant under time
evolution. This is reflected in the isomorphism of the reduced phase spaces mentioned above.
\qeb

\ebm

\section{Stratified quantum theory}
\label{A-SQM}

\subsection{Quantization and reduction}

To construct the quantum theory of the reduced system, one may either first reduce the classical system and then quantize or first quantize and then reduce the quantum system. Here, we follow the second strategy, that is, we carry out geometric (K\"ahler) quantization on $\ctg G^N$ and subsequent reduction. Let $\mf g_\CC$ denote the complexification of $\mf g$ and let $G_\CC$ denote the complexification of $G$. This is a complex Lie group having $G$ as its maximal compact subgroup. It is unique up to isomorphism. For $G = \SU(n)$, we have $G_\CC = \SL(n,\CC)$. By restriction, the exponential mapping  
$$
\exp : \mf g_\CC \to G_\CC
$$
of $G_\CC$ and multiplication in $G_\CC$ induce a diffeomorphism
\beq\label{G-poldec-1}
G\times\mf g \to G^\CC
 \,,~~~~~~
(a,A) \mapsto a\exp(\mr i A)\,,
\eeq
which is equivariant with respect to the action of $G$ on $G\times \mf g$ by 
$$
g \cdot (a,A) := \big(g a g^{-1} , \Ad(g) A\big)
$$
and the action of $G$ on $G_\CC$ by conjugation. For $G = \SU(n)$, this diffeomorphism amounts to the inverse of the polar decomposition. By applying this diffeomorphism	to each copy, we obtain a diffeomorphism
$$
G^N\times\mf g^N \to G_\CC^N
 \,,~~
(a_1,\dots,a_N , A_1,\dots,A_N)
 \mapsto 
\big(a_1\exp(\mr i A_1),\dots,a_N\exp(\mr i A_N)\big)\,.
$$
By composing the latter with the global trivialization \eqref{G-trviz-N}, we obtain a diffeomorphism 
\beq\label{G-dfm-P}
\ctg G^N \to G_\CC^N
\eeq 
which, due to \eqref{G-Wir-P}, is equivariant with respect to the lifted action of $G$ on $\ctg G^N$ and the action of $G$ on $G_\CC^N$ by diagonal conjugation. Via this diffeomorphism, the complex structure of $G_\CC^N$ and the symplectic structure of $\ctg G^N$ combine to a K\"ahler structure. Half-form K\"ahler quantization on $G_\CC^N$ yields the Hilbert space 
$$
HL^2(G_\CC^N , \mr d \nu)
$$
of holomorphic functions on $G_\CC^N$ which are square-integrable with respect to the measure 
$$
\mr d \nu = \mr e^{- \kappa / \hbar} \, \eta \, \ve\,,
$$
where 
$$
\kappa(a_1 \mr e^{\mr i A_1} , \dots , a_N \mr e^{\mr i A_N})
 =
|A_1|^2 + \cdots + |A_N|^2
$$
is the K\"ahler potential on $G_\CC^N$,
$$
\eta(a_1 \mr e^{\mr i A_1} , \dots , a_N \mr e^{\mr i A_N})
 =
\sqrt{\det\frac{\sin\big(\ad(A_1)\big)}{\ad(A_1)}}
 \cdots
\sqrt{\det\frac{\sin\big(\ad(A_N)\big)}{\ad(A_N)}}
$$
is the half-form correction and 
$$
\ve(a_1 \mr e^{\mr i A_1} , \dots , a_N \mr e^{\mr i A_N})
 = 
\mr d a_1 \cdots \mr d a_N \, \mr d A_1 \cdots \mr d A_N
$$
is the Liouville measure on $\ctg G^N$. Reduction then yields the closed subspace 
$$
\mc H = HL^2(G_\CC^N , \mr d \nu)^G
$$
of $G$-invariants as the Hilbert space of the reduced system.  

\bbm

The above result belongs to Hall \cite{Hall:cptype}. Alternatively, the Hilbert space $HL^2(G_\CC^N , \mr d \nu)$ is obtained via the Segal-Bargmann transformation for compact Lie groups \cite{Hall:SBT}.
\qeb

\ebm

\subsection{Orbit type costratification}

Following Huebsch\-mann \cite{Hue:Quantization}, we are going to define the subspaces associated with the orbit type strata of $\pha$ to be the orthogonal complements of the subspaces of functions vanishing at those strata. To follow this idea, we first have to clarify how to interpret elements of $\mc H$ as functions on $\pha$. In the case $N=1$ discussed in \cite{HRS} and \cite{adg}, this is readily done by observing that $\pha \cong T_\CC/W$, where $T$ is a maximal torus in $G$ and $W$ the corresponding Weyl group, and using the isomorphism $HL^2(G_\CC , \mr d \nu)^G \cong HL^2(T_\CC,\mr d \nu_T)^W$, see \S 3.1 of \cite{HRS}. Here, the measure $\mr d \nu_T$ is obtained from $\mr d \nu$ by integration over the conjugation orbits in $G_\CC$, thus yielding an analogue of Weyl's integration formula for $HL^2(G_\CC^N , \mr d \nu)$. In the general case, the argument is as follows. 

First, we construct a quotient of $G_\CC^N$ on which the elements of $\mc H$ define functions. Consider the action of $G_\CC$ on $G_\CC^N$ by diagonal conjugation. For $\ul a \in G_\CC^N$, let $G_\CC \cdot \ul a$ denote the corresponding orbit. Since $G_\CC$ is not compact, $G_\CC \cdot \ul a$ need not be closed. If a holomorphic function on $G_\CC^N$ is invariant under the action of $G$ by diagonal conjugation, then it is also invariant under the action of $G_\CC$ by diagonal conjugation, i.e., it is constant on the orbit $G_\CC \cdot \ul a$ for every $\ul a \in G_\CC^N$. Being continuous, it is then constant on the closure $\ol{G_\CC \cdot \ul a}$. As a consequence, it takes the same value on two orbits whenever their closures intersect. This motivates the following definition. Two elements $\ul a , \ul b \in G_\CC^N$ are said to be orbit closure equivalent if there exist $\ul c_1 , \dots , \ul c_r \in G_\CC^N$ such that
$$
\ol{G_\CC \cdot \ul a} \cap \ol{G_\CC \cdot \ul c_1} \neq \varnothing
 \,,~~~
\ol{G_\CC \cdot \ul c_1} \cap \ol{G_\CC \cdot \ul c_2} \neq \varnothing
 \,,~~~ \dots \,,~~~
\ol{G_\CC \cdot \ul c}_r \cap \ol{G_\CC \cdot \ul b} \neq \varnothing\,.
$$
Clearly, orbit closure equivalence defines an equivalence relation on $G_\CC^N$, indeed. Let $G_\CC^N // G_\CC$ denote the topological quotient\footnote{This notation is motivated by the fact that the quotient provides a categorical quotient of $G_\CC^N$ by $G_\CC$ in the sense of geometric invariant theory \cite{Mumford}.}. By construction, the elements of $\mc H$ descend to continuous functions on $G_\CC^N // G_\CC$. 

Second, we recall from the literature how the orbit closure quotient $G_\CC^N // G_\CC$ is related to the reduced phase space $\pha$. We follow \cite{HeinznerLoose}, which in our opinion is particularly transparent. Via the equivariant diffeomorphism \eqref{G-dfm-P}, we can view the momentum mapping as a mapping
$$
\mu : G_\CC^N \to \mf g^\ast
$$
and we can view $\pha$ as the quotient of $\mu^{-1}(0) \subset G_\CC^N$ by the action of $G$. 
For the situation we are interested in, we may assume $G_\CC$ to be linear algebraic. Then, $G_\CC^N$ is an affine variety in some complex vector space $V$, the action of $G$ on $G_\CC^N$ by diagonal conjugation is the restriction of a representation of $G$ on $V$ to an action of $G$ on $G_\CC^N$ and the momentum mapping is the restriction to $G_\CC^N$ of the mapping 
$$
\tilde\mu : V \to \mf g^\ast
 \,,~~~~~~
\tilde\mu(v)(A) := \frac{1}{2\mr i} \, \langle v, A v \rangle\,,
$$
where $\langle \,\cdot\, , \,\cdot\, \rangle$ is an appropriate $G$-invariant scalar product on $V$ and $A$ acts on $v$ by the induced representation of the Lie algebra. In this situation, the level set $\mu^{-1}(0)$ has the following properties \cite{KempfNess}. 

\ben

\item \label{i-KN-0} For all $\ul a\in G_\CC^N$, one has $\ol{G_\CC \cdot \ul a} \cap \mu^{-1}(0) \neq \varnothing$.

\item \label{i-KN-1} For all $\ul a\in G_\CC^N$, the orbit $G_\CC \cdot \ul a$ is closed iff $G_\CC \cdot \ul a \cap \mu^{-1}(0) \neq \varnothing$.

\item \label{i-KN-2} For all $\ul a\in \mu^{-1}(0)$, one has $G_\CC \cdot \ul a \cap \mu^{-1}(0) = G \cdot \ul a$.

\een

Properties \ref{i-KN-1} and \ref{i-KN-2} ensure that $\mu^{-1}(0)$ is what is known in geometric invariant theory as a Kempf-Ness set. Using properties \ref{i-KN-0}--\ref{i-KN-2}, one can prove the following.

\btm\label{T-KN}

The natural inclusion mapping $\mu^{-1}(0) \to G_\CC^N$ induces a homeomorphism 
\beq\label{G-KN-Hoem}
\pha \to G_\CC^N // G_\CC\,.
\eeq

\etm

\bbw

See \cite{HeinznerLoose}.
\ebw

As a by-product of the proof, one finds that two points $\ul a , \ul b \in G_\CC^N$ are orbit closure equivalent iff 
\beq\label{G-KN-rel}
\ol{G_\CC \cdot \ul a} \cap \ol{G_\CC \cdot \ul b} \cap \mu^{-1}(0) \neq \varnothing\,.
\eeq
As a result, via the homeomorphism \eqref{G-KN-Hoem}, the elements of $\mc H$ can be interpreted as functions on $\pha$. By virtue of this interpretation, to a given orbit type stratum $\pha_\tau \subseteq \pha$, there corresponds the closed subspace 
$$
\mc V_{\tau} := \{\psi \in \mc H : \psi_{\res \pha_\tau} = 0\}\,.
$$
We define the subspace $\mc H_\tau$ associated with $\pha_\tau$ to be the orthogonal complement of $\mc V_\tau$ in $\mc H$. Then, we have the orthogonal decomposition
$$
\mc H_\tau \oplus \mc V_\tau = \mc H\,.
$$

\bbm\label{Bem-HS}

Since holomorphic functions are continuous, one has 
\beq\label{G-D-V}
\mc V_{\tau} := \{\psi \in \mc H : \psi_{\res \ol{\pha_\tau}} = 0\}\,.
\eeq
First, since the principal stratum is dense in $\pha$, this implies that the
subspace associated with that stratum coincides with $\mc H$. Thus, in the
discussion of the orbit type subspaces below, the principal stratum may be
ignored. Second, recall that in a stratification, the strata satisfy the
condition of the frontier, which means that if $\pha_{\sigma} \cap
\ol{\pha_{\tau}} \neq \varnothing$, then $\pha_{\sigma} \subset \ol{\pha_{\tau}}$.
In view of this, \eqref{G-D-V} implies that if $\pha_{\sigma} \cap
\ol{\pha_{\tau}} \neq \varnothing$, then $\mc V_\tau \subseteq \mc V_\sigma$ and
hence $\mc H_\sigma \subseteq \mc H_\tau$. The family of orthogonal projections  
$$
\mc H_\tau \to \mc H_\sigma
 \text{ whenever } 
\pha_{\sigma} \cap \ol{\pha_{\tau}} \neq \varnothing
$$
makes the family of closed subspaces $\mc H_\tau$ into a costratification in the
sense of Huebsch\-mann \cite{Hue:Quantization}. 
\qeb

\ebm

In order to analyse the condition $\psi_{\res \pha_\tau} = 0$, it is convenient to work with the subsets of $G_\CC^N$ which under the natural projection $G_\CC^N \to G_\CC^N // G_\CC$ and the homeomorphism \eqref{G-KN-Hoem} correspond to the orbit type strata of $\pha$. For a given orbit type stratum $\pha_\tau$, denote this subset by $(G_\CC^N)_\tau$. That is, $(G_\CC^N)_\tau$ consists of the elements $\ul a$ of $G_\CC^N$ whose orbit closure equivalence class belongs to the image of $\pha_\tau$ under the homeomorphism \eqref{G-KN-Hoem}. In other words, $\ul a \in (G_\CC^N)_\tau$ iff it is orbit closure equivalent to some element of $\mu^{-1}(0)$ whose $G$-orbit belongs to $\pha_\tau$. Clearly,
\beq\label{G-V-GC}
\mc V_\tau = \{\psi \in \mc H : \psi_{\res (G_\CC^N)_\tau} = 0\}\,.
\eeq

 \comment{
\bsz\label{S-strata}

Let $\pha_\tau \subseteq \pha$ be an orbit type stratum and let $\ul a \in G_\CC^N$. Then, $\ul a \in (G_\CC^N)_\tau$ iff $\ol{G_\CC \cdot \ul a}$ contains a point of $\mu^{-1}(0)$ whose $G$-orbit belongs to $\pha_\tau$.

\esz

\bbw

Clearly, $\ul a \in (G_\CC^N)_\tau$ iff $\ul a$ is orbit closure equivalent to some element $b$ of $\mu^{-1}(0)$ whose $G$-orbit belongs to $\pha_\tau$. We show that if $\ul b\in\mu^{-1}(0)$, then $\ul a$ and $\ul b$ are orbit closure equivalent iff $\ul b \in \ol{G_\CC \cdot \ul a}$. As noted above, they are orbit closure equivalent iff condition \eqref{G-KN-rel} holds. If $b \in \mu^{-1}(0)$, property \ref{i-KN-2} above implies that $G_\CC \cdot \ul b \cap \mu^{-1}(0) = G \cdot \ul b$ and hence $\ol{G_\CC \cdot \ul b} \cap \mu^{-1}(0) = G \cdot \ul b$, because both $\mu^{-1}(0)$ and $G \cdot \ul b$ are closed. Therefore, if $\ul b \in \mu^{-1}(0)$, then condition \eqref{G-KN-rel} reads 
$$
\ol{G_\CC \cdot \ul a} \cap (G \cdot \ul b) \neq \varnothing\,.
$$
Since $\ol{G_\CC \cdot \ul a}$ is invariant under $G$, this is equivalent to $\ul b \in \ol{G_\CC \cdot \ul a}$.
\ebw
 }

\subsection{Characterization of costrata in terms of relations}
\label{A-SQT-AA-defRel}

To conclude the general discussion, we describe how to construct $\mc V_\tau$ and $\mc H_\tau$ using defining relations for the orbit type strata $\pha_\tau$.

Let $\mc R$ denote the algebra of $G$-invariant representative functions on $G^N_\CC$. Since $G_\CC^N$ is the complexification of the compact Lie group $G^N$, Theorem 3 in \cite{Procesi:LG} implies that $\mc R$ coincides with the coordinate ring on $G_\CC^N$. Recall that an ideal $\mc I \subset \mc R$ is called a radical ideal if for all $f \in \mc R$ satisfying $f^n \in \mc I$ for some $n$ one has $f \in \mc I$. Moreover, given a subset $A \subset \mc R$, one defines the zero locus of $A$ by 
$$
N := \{\ul a \in G^N_\CC : f(\ul a) = 0 \text{ for all } f \in A\} \subset G^N_\CC\,.
$$
It coincides with the zero locus of the ideal in $\mc R$ generated by $A$. Conversely, given a subset $B \subset G^N_\CC$, one defines the vanishing ideal of $B$ in $\mc R$ by 
$$
\mc V(B) := \{f \in \mc R : f_{\res B} = 0 \}\,.
$$
By analogy, one defines the vanishing ideal $\mc V_{\text{hol}}(B)$ of $B$ in the ambient algebra $\text{Hol}(G^N_\CC)^G$.

\bsz\label{S-V}

Let $\pha_\tau$ be an orbit type stratum and let $A_\tau$ be a subset of $\mc R$ satisfying

\ben

\sitem\label{i-S-V-Loc}
The zero locus of $A_\tau$ coincides with the topological closure of $(G_\CC^N)_\tau$,

\sitem\label{i-S-V-Rad}
The ideal generated by $A_\tau$ in $\mc R$ is a radical ideal.

\een

Then, $\mc V_\tau$ is obtained by intersecting $\mc H$ with the ideal generated algebraically by $A_\tau$ in the algebra $\text{\rm Hol}(G^N_\CC)^G$ of $G$-invariant holomorphic functions on $G_\CC^N$.

\esz

\bpf

Let $N_\tau \subset G_\CC^N$ denote the zero locus of $\mc A_\tau$. On the one hand, by Hilbert's Nullstellensatz, condition \rref{i-S-V-Rad} implies that $\mc V(N_\tau)$ coincides with the ideal generated by $A_\tau$ in $\mc R$ and hence that $\mc V(N_\tau)$ is generated by $A_\tau$. Then, by Proposition 4 in \cite{Serre}, $\mc V_{\text{hol}}(N_\tau)$ is generated algebraically by $A_\tau$, too. By condition \rref{i-S-V-Loc}, we can replace $N_\tau$ by the topological closure of $(G_\CC^N)_\tau$ and, consequently, by $(G_\CC^N)_\tau$. Since $\mc V_\tau = \mc V_{\text{hol}}\big((G_\CC^N)_\tau\big) \cap \mc H$, the assertion follows. 
\epf

By Hilbert's Basissatz, finite subsets $A_\tau \subset \mc R$ satisfying conditions \rref{i-S-V-Loc} and \rref{i-S-V-Rad} of Proposition \rref{S-V} exist. Below, we will derive $A_\tau$ for a particular stratum in the case $G = \SU(2)$. Given $A_\tau$, Proposition \rref{S-V} implies the following explicit characterization of the subspaces $\mc V_\tau$ and $\mc H_\tau$ in terms of multiplication operators. For $f \in \mc R$, let $\hat f : \mc H \to \mc H$ denote the operator of multiplication by $f$.

\bfg\label{F-V}

Let $\pha_\tau$ be an orbit type stratum and let $A_\tau = \{p_1 , \dots , p_r\}$ be a finite subset of $\mc R$ satisfying conditions \rref{i-S-V-Loc} and \rref{i-S-V-Rad} of Proposition \rref{S-V}. Then,
\eqqedan
\beq
\mc V_\tau = \im(\hat p_1) + \cdots + \im(\hat p_r)
 \,,\qquad
\mc H_\tau = \ker\big(\hat p_1^\dagger\big) \cap \cdots \cap \ker\big(\hat p_r^\dagger\big)\,.
\eqqed
\eeq
\eqqedaus

\efg

In what follows, we will refer to conditions \rref{i-S-V-Loc} and \rref{i-S-V-Rad} of Proposition \rref{S-V} as the zero locus condition and the radical ideal condition, respectively.

\section{Orbit type strata}

From now on, let $G = \SU(2)$. Then, $G_\CC = \SL(2,\CC)$, $\mf g = \su(2)$ and $\mf g = \sl(2,\CC)$. For convenience, we keep the notation $G$ and $G_\CC$. We are going to characterize the subsets $(G_\CC^N)_\tau$ of $G_\CC^N$ corresponding to the orbit type strata $\pha_\tau$ of $\pha$. First, we determine the orbit types of the lifted action of $G$ on $\ctg G^N$. For that purpose, we use the global trivialization \eqref{G-trviz-N} to pass to $G^N \times \mf g^N$ with the action of $G$ given by \eqref{G-Wir-P}. The stabilizer of an element $(\ul a , \ul A) \in G^N \times \mf g^N$ is given by 
$$
\mr C_G(a_1) \cap \cdots \cap \mr C_G(a_N) \cap \mr C_G(A_1) \cap \cdots \cap \mr C_G(A_N)\,,
$$
where $\mr C_G(\cdot)$ denotes the respective centralizer in $G$, i.e., 
$$
\mr C_G(a_i) = \{g \in G : g a_i g^{-1} = a_i\}
 \,,~~~~~~
\mr C_G(A_i) = \{g \in G : \Ad(g) A_i = A_i\}\,.
$$
Let $Z$ denote the center of $G$. This is also the center of $G_\CC$. Let $T \subseteq G$ denote the subgroup of diagonal matrices. Clearly, $T$ is a maximal toral subgroup, isomorphic to $\mr U(1)$. The centralizer $\mr C_G(a_i)$ is conjugate to $T$ unless $a_i = \pm\II$ and $\mr C_G(\pm\II) = G$. Similarly, the centralizer $\mr C_G(A_i)$ is conjugate to $T$ unless $A_i = 0$ and $\mr C_G(0) = G$. Since two distinct subgroups which both are conjugate to $T$ intersect in $Z$, by taking intersections, we see that the stabilizer can be $G$, conjugate to $T$, or $Z$, where $Z$ is the generic situation. Accordingly, there are three orbit types and these can be labeled by $G$, $T$ and $Z$, where $Z$ is the principal orbit type. We describe the corresponding orbit type subsets of $G^N \times \mf g^N$:

\ben

\item[$(G)$] For all $i$, one must have $\mr C_G(a_i) = \mr C_G(A_i) = G$.
Hence, $(\ul a , \ul A)$ has orbit type $G$ iff 
$$
(\ul a , \ul A) \in Z^N \times \{0\}^N\,.
$$

\item[$(T)$] Up to conjugacy, one of the centralizers $\mr C_G(a_i)$ or $\mr
C_G(A_i)$ must be equal to $T$ and all the other centralizers must contain $T$.
If $\mr C_G(a_j)$ contains $T$, then $a_j \in T$. If $\mr C_G(A_j)$ contains
$T$, then $A_j \in \mf t$, the Lie subalgebra of $\mf g$ corresponding to $T$,
i.e., the Lie subalgebra of diagonal matrices in $\mf g$. Hence, $(\ul a , \ul
A)$ has orbit type $T$ iff it is conjugate to an element of the subset 
$$
\left(T^N \times \mf t^N\right) \setminus \left(Z^N \times \{0\}^N\right)\,.
$$

\item[$(Z)$] Clearly, $(\ul a , \ul A)$ has orbit type $Z$ iff it does not have
obit type $T$ or $G$, i.e., iff it is not conjugate to an element of $T^N \times
\mf t^N$.  

\een

Next, we intersect the orbit type subsets with the momentum level set
$\mu^{-1}(0)$. According to \eqref{G-ImpAbb}, 
$$
T^N \times \mf t^N \subset \mu^{-1}(0)\,.
$$
Since $\mu^{-1}(0)$ is $G$-invariant, this implies that the subsets of orbit type $T$ and orbit type $G$ are contained in $\mu^{-1}(0)$. For $N=1$, the condition $\mu(a,A) = 0$ implies that $a$ and $A$ commute and hence that they are simultaneously diagonalizable. Hence, in this case, in $\pha$ only the orbit types $G$ and $T$ occur. Since this case has been discussed in detail in \cite{HRS}, in what follows we may restrict attention to the case $N \geq 2$. Here, the subsets of orbit type $G$ and $T$ do not exhaust $\mu^{-1}(0)$. Therefore, all three orbit types survive the reduction procedure, thus yielding three orbit type subsets of $\pha$. To find the orbit type strata, we have to decompose these orbit type subsets into connected components.

\ben

\item[$(G)$] Since the elements of $Z^N \times \{0\}^N$ are invariant under the action of $G$, each of them projects to a single point in $\pha$. Therefore, there exist $2^N$ orbit type strata of orbit type $G$, each of which consists of a single point representing the (trivial) orbit  of an element of $Z^N \times \{0\}^N$. Since such an element is of the form $(\nu_1 \II , \dots , \nu_N \II , 0, \dots , 0)$ for some sequence of signs $\ul\nu = (\nu_1 , \dots , \nu_N)$, we denote the corresponding stratum by $\pha_{\ul\nu}$. 

\item[$(T)$] Since $Z^N \times \{0\}^N$ consists of finitely many points and
$T^N \times \mf t^N$ has dimension at least $2$, the complement $(T^N \times \mf
t^N) \setminus (Z^N \times \{0\}^N)$ is connected. Since the subset of $\pha$ of
orbit type $T$ is the image of $(T^N \times \mf t^N) \setminus (Z^N \times
\{0\}^N)$ under the natural projection $\mu^{-1}(0) \to \pha$, it is connected,
too. Hence, it forms an orbit type stratum. We denote this stratum by $\pha_T$.

\item[$(Z)$] Since $\mf g^\ast$ has dimension $3$, the level set $\mu^{-1}(0)$ generically has dimension $2N \cdot 3 - 3 = 3(2N-1)$. On the other hand, since $T$ has dimension $1$ and the elements of $T^N \times \mf t^N$ have stabilizer $T$ under the action of $G$, the subset of $G^N \times \mf g^N$ of orbit type $T$ has dimension $2N \cdot 1 + (3-1) = 2(N+1)$. Hence, if the orbit type $Z$ occurs in $\pha$, i.e., if $N \geq 2$, then the subset of $\mu^{-1}(0)$ generated from $T^N \times \mf t^N$ by the action of $G$ has codimension
$$
3(2N-1) - 2(N+1) = 4N-5 \geq 3\,.
$$
Therefore, its complement is connected. Since the complement coincides with the subset of $\mu^{-1}(0)$ of orbit type $Z$, the subset of $\pha$ of this orbit type is connected. Hence, it forms an orbit type stratum. We denote this stratum by $\pha_Z$.

\een

We can visualize the set of strata, together with its natural partial ordering defined by 
$$
\tau \leq \tau' ~ \bem{ iff } ~ \pha_\tau \subset \ol{\pha_{\tau'}}\,,
$$
in a Hasse diagram, where a line running from $\tau$ on the left to $\tau'$ on the right means that $\tau \leq \tau'$:
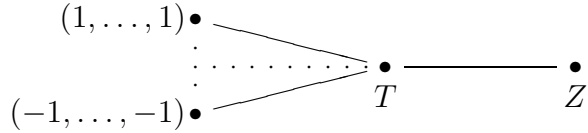
\begin{center}
\unitlength2.5cm
\begin{picture}(2,1.5)
\put(0,0.75){
%
\plvmee{0,0.25}{cr}{(1 , \dots , 1)}
\vpunkte{0,0}
\pslene{0,0}
\plvee{0,-0.25}{cr}{(-1 , \dots , -1)}
%
\plene{1,0}{tc}{T}
\whole{2,0}{tc}{Z}
}
\end{picture}
\end{center}
Finally, we transport these results to $G_\CC^N$, that is, for each of the strata $\tau$ just found, we characterize the subset $(G_\CC^N)_\tau$ of $G_\CC^N$. It suffices to do this for every sequence of signs $\ul\nu = (\nu_1 , \dots , \nu_N)$ and for $T$. Let $T_\CC \subset G_\CC$ denote the subgroup of diagonal matrices.

\btm\label{T-OT}

Let $\ul a \in G_\CC^N$. Then,

\ben

\sitem $\ul a \in (G_\CC^N)_{\ul \nu}$ iff $\ul a$ is orbit closure equivalent to $(\nu_1 \II , \dots , \nu_N \II)$,

\sitem $\ul a \in (G_\CC^N)_T$ iff $\ul a$ is orbit closure equivalent to an element of $T_\CC^N \setminus Z^N$.

\een

\etm

\bbw

1.\abs By definition, $\ul a$ belongs to $(G_\CC^N)_{\ul \nu}$ iff it is orbit closure equivalent to an element $\ul b$ of $\mu^{-1}(0)$ whose $G$-orbit belongs to the stratum $\pha_{\ul \nu}$. As we have seen above, the latter holds iff $\ul b$ is the image of the point $\big((\nu_1 \II , \dots , \nu_N \II) , (0 , \dots , 0)\big)$ under the diffeomorphism \eqref{G-dfm-P}, that is, iff
$
\ul b = (\nu_1 \II , \dots , \nu_N \II)\,.
$

2.\abs Similarly, by definition, $\ul a$ belongs to $(G_\CC^N)_T$ iff it is
orbit closure equivalent to an element $\ul b$ of $\mu^{-1}(0)$ whose $G$-orbit
belongs to the stratum $\pha_T$. By the discussion above, the latter holds iff
the preimage of $\ul b$ under the diffeomorphism \eqref{G-dfm-P} is conjugate
under $G$ to a point of $(T^N \times \mf t^N) \setminus (Z^N \times \{0\}^N)$.
Since the diffeomorphism \eqref{G-dfm-P} is $G$-equivariant, this condition is
equivalent to the condition that $\ul b$ be conjugate under $G$ to a point in
the image of $(T^N \times \mf t^N) \setminus (Z^N \times \{0\}^N)$ under
\eqref{G-dfm-P}, i.e., to a point in $T_\CC^N \setminus Z^N$. Since two points
of $\mu^{-1}(0)$ are orbit closure equivalent iff they are conjugate under $G$,
it follows that $\ul a$ belongs to $(G_\CC^N)_T$ iff it is orbit closure
equivalent to an element of $T_\CC^N \setminus Z^N$. 
\ebw

\section{Zero locus condition}

In this section, for the strata $\tau$ found above, we determine finite subsets $A_\tau$ of $\mc R$ having the corresponding orbit type subset $(G_\CC^N)_\tau$ as their zero locus. Since $\tau = Z$ correponds to the principal stratum and hence, by Remark \rref{Bem-HS}, $\mc H_Z = \mc H$, it suffices to discuss the the secondary strata $\tau = \ul \nu$ and $\tau = T$. First, consider the stratum $T$.

\btm\label{T-Rel-T}

The topological closure $\ol{(G_\CC^N)_T}$ is the set of common zeros of the $G$-invariant representative functions
$$
p^T_{ij}(\ul a) := \tr\big([a_i,a_j]^2\big) \,,\qquad 1 \leq i < j \leq N 
\phantom{< k\,.}
$$
and
$$
p^T_{ijk}(\ul a) := \tr\big([a_i,a_j]a_k\big) \,,\qquad 1 \leq i < j < k \leq N\,.
$$

\etm

\bbw

By Theorem \rref{T-OT}, we have to show that $\ul a$ is orbit closure equivalent to an element of $T_\CC^N$ iff 
\beq\label{G-T-Rel-T-1}
p^T_{ij}(\ul a) = 0 \,,\qquad 1 \leq i < j \leq N \phantom{< k\,.}
\eeq
and
\beq\label{G-T-Rel-T-2}
p^T_{ijk}(\ul a) = 0 \,,\qquad 1 \leq i < j < k \leq N\,.
\eeq
First, assume that $\ul a$ is orbit closure equivalent to some $\ul b \in T_\CC^N$. Then, $\psi(\ul a) = \psi(\ul b)$ for any continuous invariant function $\psi$. Hence, $p^T_{ij}(\ul a) = p^T_{ij}(\ul b) = 0$ and $p^T_{ijk}(\ul a) = p^T_{ijk}(\ul b) = 0$, because the members of $\ul b$ commute pairwise. Hence, the conditions \eqref{G-T-Rel-T-1} and \eqref{G-T-Rel-T-2} hold for $\ul a$. 

Now, conversely, assume that $\ul a$ satisfies the conditions \eqref{G-T-Rel-T-1} and \eqref{G-T-Rel-T-2}. If $\ul a \in Z^n$, we are done. Otherwise, there is a smallest $i$ such that $a_i \notin Z$. There exists $g \in G_\CC$ such that $g a_i g^{-1}$ has Jordan normal form. Since $a_1, \dots , a_{i-1} \in Z$, we have
$$
g \cdot \ul a = \left(a_1 , \dots , a_{i-1} , g a_i g^{-1} , \dots , g a_N g^{-1}\right)\,.
$$
This shows that up to the action of $G_\CC$ we may assume that the first noncentral entry $a_i$ has Jordan normal form. Then, the following two cases can occur.

\ben

\item[] Case (a):\abs $a_i = \bbma \alpha & 0 \\ 0 & \frac 1 \alpha \ebma$~ with $\alpha \in \CC$, $\alpha \neq 0 , \pm1$.

\item[] Case (b):\abs $a_i = \bbma \alpha & 1 \\ 0 & \alpha \ebma$~ with $\alpha = \pm1$.

\een

Writing 
$$
a_j = \bbma \beta_{11} & \beta_{12} \\ \beta_{21} & \beta_{22} \ebma\,,
$$
we compute
\beq\label{G-T-Rel-T-3}
p^T_{ij}(\ul a)
 =
 \bca
- 2 \left(\alpha - \frac 1 \alpha\right)^2 \beta_{12} \, \beta_{21} & \bem{case (a),}
\\
2 \beta_{21}^2 & \bem{case (b).}
 \eca
\eeq
In case (b), it follows that the matrices $a_{i+1} , \dots , a_N$ are upper triangular. Hence, all the matrices $a_1 , \dots , a_N$ are upper triangular. Since for a triangular matrix and $n \in \NN$ one has 
$$
\bbma \frac 1 n & 0 \\ 0 & n \ebma 
\bbma \beta & \gamma \\ 0 & \frac 1 \beta \ebma
\bbma n & 0 \\ 0 & \frac 1 n \ebma 
 =
\bbma \beta & \frac{1}{n^2} \, \gamma \\ 0 & \frac 1 \beta \ebma\,,
$$
in this case the sequence 
$$
\bbma \frac 1 n & 0 \\ 0 & n \ebma \cdot \ul a
 \,,\quad
n \in \NN\,,
$$
converges to an element of $T_\CC^N$. Consequently, $\ul a$ is orbit closure equivalent to an element of $T_\CC^N$. 

In case (a), on the other hand, \eqref{G-T-Rel-T-3} implies that the matrices $a_{i+1} , \dots , a_N$ are triangular, but it does not tell us whether they are upper or lower triangular. In fact, there exist elements $\ul a$ of $G_\CC^N$ which satisfy \eqref{G-T-Rel-T-1} and which contain both types of triangular matrices, see the remark below.
Hence, in case (a), we have to take into account the conditions \eqref{G-T-Rel-T-2}. We show that these conditions imply that all entries of $\ul a$ are triangular of the same type. Assume, on the contrary, that $a_j$ is upper triangular and that $a_k$ is lower triangular. Writing
$$
a_j = \bbma \beta & \gamma \\ 0 & \frac 1 \beta \ebma
 \,,~~~~~~
a_k = \bbma \delta & 0 \\ \ve & \frac 1 \delta \ebma\,,
$$
where $\gamma, \ve \neq 0$, we compute
$$
p^T_{ijk}(\ul a)
 =
\left(\alpha - \frac 1 \alpha\right) \gamma \, \ve \neq 0
$$
(contradiction). Thus, all the $a_i$ are triangular of the same type. Then, the same argument as in case (b) shows that $\ul a$ is orbit closure equivalent to an element of $T_\CC^N$.
\ebw

\bbm~

\ben

\item The conditions \eqref{G-T-Rel-T-1} cannot exclude the situation that $\ul a$ contains triangular matrices of different types. To see this,  assume, for example, that $\ul a$ contains 
$$
a_i = \bbma \alpha & 0 \\ 0 & \frac 1 \alpha \ebma 
 \,,~~~~~~
a_j = \bbma \beta & \gamma \\ 0 & \frac 1 \beta \ebma 
 \,,~~~~~~
a_k = \bbma \delta & 0 \\ \ve & \frac 1 \delta \ebma
$$
with $\gamma , \ve \neq 0$. Then, as shown in the proof,
$$
p^T_{ij}(\ul a) = p^T_{ik}(\ul a) = 0\,.
$$
On the other hand, we compute
$$
p^T_{jk}(\ul a)
 = 
2 \gamma \ve 
 \left(
\gamma \ve + \left(\beta - \frac 1 \beta\right) \left(\delta - \frac 1 \delta\right)
 \right)\,.
$$
Hence, $p^T_{jk}(\ul a) = 0$ for arbitrary values of $\beta , \gamma , \delta$ and 
$$
\ve
 = 
- \frac 1 \gamma \left(\beta - \frac 1 \beta\right) \left(\delta - \frac 1 \delta\right)\,.
$$

\item We show that the functions $p^T_{ij}$ can be rewritten as 
\beq\label{G-Flo}
p^T_{ij}(\ul a)
 =
2
 \left(
\tr(a_i a_j)^2
 -
\tr(a_i) \tr(a_j) \tr(a_i a_j)
 +
\tr(a_i)^2
 +
\tr(a_j)^2
 -
4
 \right)\,.
\eeq
According to the Cayley-Hamilton theorem, every complex $(2\times 2)$-matrix $a$ satisfies the relation
$$
\chi_a(a) = 0\,,
$$
where 
$$
\chi_a(z) = \det(z\II - a)
$$
is the characteristic polynomial of $a$. Evaluation of the determinant yields 
$$
\chi_a(z) = z^2 - \tr(a) z + \det(a)\,.
$$
Hence, every $a \in G_\CC = \SL(2,\CC)$ satisfies the relation
\beq\label{G-CHT}
a^2 = \tr(a) a - \II.
\eeq
Now, \eqref{G-Flo} follows by writing 
$$
p^T_{ij}(\ul a)
 =
\tr\big([a_i , a_j]^2\big)
 =
2 \tr\big((a_i a_j)^2\big) - 2 \tr\big(a_i^2 a_j^2\big)
$$
and replacing all squares on the right hand side according to \eqref{G-CHT}.
\qeb

\een

\ebm

Now, we turn to the discussion  of the strata labeled by sequences of signs $\ul \nu$. By Theorem \rref{T-OT}, $(G_\CC^N)_{\ul \nu}$ is the orbit closure equivalence class of the single point $(\nu_1 \II , \dots , \nu_N \II)$. Hence, it is closed.

\btm\label{T-Rel-G}

The subset $(G_\CC^N)_{\ul \nu} \subset G_\CC^N$ is the set of common zeros of the $G_\CC$-invariant functions $p^T_{ij}$, $1 \leq i < j \leq N$, $p^T_{ijk}$, $1 \leq i < j < k \leq N$, and 
$$
p^{\ul \nu}_i(\ul a) := \tr(a_i) - \nu_i 2
 \,,\qquad
i = 1 , \dots , N\,.
$$

\etm

\bbw

First, assume that $\ul a$ belongs to $(G_\CC^N)_{\ul \nu}$. Then, it is orbit closure equivalent to $\ul b = (\nu_1 \II , \dots , \nu_N \II)$ and hence
$$
p^T_{ij}(\ul a) = p^T_{ij}(\ul b) = \tr\big([\nu_i \II , \nu_j \II]^2\big) = 0
$$
for all $i$, $j$, 
$$
p^T_{ijk}(\ul a) = p^T_{ijk}(\ul b) = \tr\big([\nu_i \II , \nu_j \II] \nu_k \II\big) = 0
$$
for all $i$, $j$, $k$ and
$$
p^{\ul \nu}_i(\ul a) = p^{\ul \nu}_i(\ul b) = \tr(\nu_i \II) - \nu_i 2 = 0
$$
for all $i$. Conversely, assume that $\ul a$ is a common zero of the functions
$p^T_{ij}$, $p^T_{ijk}$ and $p^{\ul \nu}_i$. As we have seen in the proof of
Theorem \rref{T-Rel-T}, the first two imply that $a_1 , \dots , a_N$ are triangular of the same type. Up to the action of $G_\CC$, we may assume that they are upper triangular, i.e.,
$$
a_i = \bbma \alpha_i & \beta_i \\ 0 & \frac{1}{\alpha_i} \ebma\,.
$$
Then,
$$
p^{\ul \nu}_i(\ul a) = \alpha_i + \frac{1}{\alpha_i} - \nu_i 2 = 0
$$
implies 
$$
(\alpha_i - \nu_i)^2
 = 
0
$$
and hence $\alpha_i = \nu_i$. Thus,
$$
a_i = \bbma \nu_i & \beta_i \\ 0 & \nu_i \ebma
$$
and by the same argument as in the proof of Theorem \rref{T-Rel-T} we can conclude that $\ul a$ is orbit closure equivalent to $(\nu_1 \II , \dots , \nu_N \II)$. 
\ebw

\bbm\label{Bem-OTnu}
 
Theorem \rref{T-Rel-G} was stated for completeness only. It is not necessary for
constructing the subspace $\mc H_{\ul \nu}$ associated with the stratum
$\pha_{\ul \nu}$. Rather, this subspace can be constructed directly as follows.
Let $\{\psi_\alpha : \alpha \in A\}$ be an orthonormal basis of $\mc H$ which
contains a constant function $\psi_0$. Such a basis exists, because the constant
functions belong to $HL^2(G_\CC^N,\mr d\nu)$ and they are invariant. Since for a
continuous invariant function $\psi$, the condition to vanish on $(G_\CC^N)_{\ul
\nu}$ is equivalent to the condition $\psi(\nu_1 \II , \dots , \nu_N \II) = 0$,
the vanishing subspace $\mc V_{\ul \nu}$ of the stratum $\pha_{\ul \nu}$, given
by \eqref{G-V-GC}, is spanned by the elements 
$$
\psi_\alpha - \psi_\alpha(\nu_1 \II , \dots , \nu_N \II) \, 1
 \,,~~~~~~
\alpha \in A\,, ~ \alpha \neq 0\,,
$$
where $1$ denotes the constant function with value $1$. 
We claim that $\mc H_{\ul \nu}$ is spanned by the single element
$$
\psi_{\ul \nu} 
 =
\frac{1}{C_{\ul \nu}} \, 
\sum_{\beta \in A}
\ol{\psi_\beta(\nu_1 \II , \dots , \nu_N \II)} \, \psi_\beta\,,
$$
where $C_{\ul \nu}$ is a normalization constant. Indeed, for any $\alpha \in A$, denoting the scalar product in $\mc H$ by $\langle \cdot , \cdot \rangle$ and writing $\ul b = (\nu_1 \II , \dots , \nu_N \II)$, we compute
 \ala{
\langle \psi_{\ul \nu} , \psi_\alpha - \psi_\alpha(\ul b) \rangle
 = &
\frac{1}{C_{\ul \nu}} \, \left(
\sum_{\beta \in A}
\psi_\beta(\ul b)
\langle \psi_\beta , \psi_\alpha \rangle
 -
\sum_{\beta \in A}
\psi_\beta(\ul b)
\psi_\alpha(\ul b)
\langle \psi_\beta , 1 \rangle
 \right).
 }
Since the basis is orthonormal, the first sum yields $\psi_\alpha(\ul b)$. Moreover, since $\psi_0$ is constant, $\langle \psi_\beta , 1 \rangle = 0$ unless $\beta = 0$. Hence, the second sum reduces to 
$$
\psi_0(\ul b)
\psi_\alpha(\ul b)
\langle \psi_0 , 1 \rangle
 =
\psi_\alpha(\ul b)
\langle \psi_0 , \psi_0(\ul b) 1 \rangle\,.
$$
Since $\psi_0(\ul b) \, 1 = \psi_0$, the scalar product gives $1$. Hence, 
$$
\langle \psi_{\ul \nu} , \psi_\alpha - \psi_\alpha(\ul b) \rangle = 0
$$
for all $\alpha \in A$, as asserted.
\qeb

\ebm

\section{Radical ideal condition}
\label{A-RadIdl}

By Remark \rref{Bem-OTnu}, checking the radical ideal condition is relevant for the stratum $\pha_T$ only. Thus, this section is devoted to the proof of the following theorem. As before, let $\mc R$ denote the algebra of $G$-invariant representative functions on $G_\CC^N$.

\btm\label{T-RI}

The ideal generated in $\mc R$ by the functions 
\beq\label{G-FnT}
p^T_{ij} \,,~~ 1 \leq i < j \leq N
 \,,\qquad
p^T_{ijk} \,,~~ 1 \leq i < j < k \leq N\,,
\eeq
is a radical ideal.

\etm

As an immediate consequence, the subspaces $\mc V_T$ and $\mc H_T$ can be characterized in terms of the multiplication operators $\hat p^T_{ij}$ and $\hat p^T_{ijk}$ as described in Corollary \rref{F-V}.
\bigskip

Denote the ideal generated in $\mc R$ by the functions \eqref{G-FnT} by $\mc I$. Let $f \in \mc R$. We have to show that if $f^n \in \mc I$ for some positive integer $n$, then $f \in \mc I$. It suffices to consider the case where $f^2 \in \mc I$, because for $n \geq 1$ the condition $f^n \in \mc I$ implies $f^{2^n} \in \mc I$, as $\mc I$ is an ideal. 

We will proceed as follows. First, we construct an adapted basis $B$ in $\mc R$ such that a subset $B_{\mc I}$ of this basis spans $\mc I$. Then, $B \setminus B_{\mc I}$ spans a vector space complement $\mc X$ of $\mc I$ in $\mc R$ and every element $h$ of $\mc R$ has a unique decomposition $h = h_{\mc I} + h_{\mc X}$ with $h_{\mc I} \in \mc I$ and $h_{\mc X} \in \mc X$. Using this decomposition, we can write
$$
f^2 = f_{\mc I}^2 + 2 f_{\mc I} f_{\mc X} + \big(f_{\mc X}^2\big)_{\mc I} + \big(f_{\mc X}^2\big)_{\mc X}
$$
to see that $f^2 \in \mc I$ implies 
\beq\label{G-Bed}
\big(f_{\mc X}^2\big)_{\mc X} = 0\,.
\eeq
The main part of the proof then consists in showing that \eqref{G-Bed} entails $f_{\mc X} = 0$. For that purpose, we will derive an approximate multiplication formula for the elements of $B \setminus B_{\mc I}$ and sort the coefficients of $f_{\mc X}$ relative to the adapted basis successively by what will be called the degree.

\subsection{Adapted basis}

\newcommand{\Sww}{\Sigma}
\newcommand{\Sws}{\ol\Sigma}
\newcommand{\Ss}{\widehat\Sigma}
\newcommand{\Sss}{\widehat{\ol\Sigma}}

For $i , j , k = 1 , \dots , N$, we define elements $t_i$, $t_{ij}$ and $t_{ijk}$ of $\mc R$ by
$$
t_i(\ul a)
 := 
\tr(a_i)
 \,,\qquad
t_{ij}(\ul a)
 := 
\tr(a_i a_j)
 \,,\qquad
t_{ijk}(\ul a)
 := 
\tr(a_i a_j a_k)\,,
$$
where $\ul a \in G^N_\CC$. According to \eqref{G-Flo},
\beq\label{G-pij}
p^T_{ij} 
 =
2 \big(t_{ij}^2 - t_i t_j t_{ij} + t_i^2 + t_j^2 - 4\big)\,,
\eeq
Moreover, using the fundamental trace identity \cite{Procesi}, which states that
$$
\tr(abc) + \tr(acb)
- \tr(ab)\tr(c) - \tr(ac)\tr(b) - \tr(bc)\tr(a)
+ \tr(a) \tr(b) \tr(c)
$$
vanishes for all two-dimensional square matrices $a,b,c$, one can check that
\beq\label{G-pijk}
p^T_{ijk} = 2 t_{ijk} - t_{ij} t_k - t_{ik} t_j - t_{jk} t_i + t_i t_j t_k\,.
\eeq
In what follows, whenever speaking of an ordering of tuples of positive integers, we mean the lexicographic ordering. For a positive integer $l$, let $\Sigma_l$ denote the set of weakly increasing finite sequences, including the trivial sequence $\varnothing$, of strongly increasing $l$-tuples of the numbers $1 , \dots , N$. Clearly, $\Sigma_1$ is just the set of weakly increasing sequences of these numbers. For any two elements $K_1 , K_2 \in \Sigma_l$, let $K_1 \sqcup K_2$ denote the element of $\Sigma_l$ obtained by concatenation of $K_1$ and $K_2$ and subsequent reordering. For $l=2$, we will also need the subset $\widehat\Sigma_2 \subset \Sigma_2$ of strongly increasing sequences.

\ble\label{L-S2}

Every $K \in \Sigma_2$ can be decomposed as $K = \hat K \sqcup \check K \sqcup \check K$ with unique $\hat K \in \widehat\Sigma_2$ and $\check K \in \Sigma_2$. 

\ele

\bpf

Assume that $K$ consists of $n_1$ pairs $(k_1,l_1)$, $n_2$ pairs $(k_2,l_2)$, etc.\ . If $n_i$ is odd, put one pair $(k_i,l_i)$ into $\hat K$ and $(n_i-1)/2$ pairs $(k_i,l_i)$ into $\check K$. If $n_i$ is even, put $n_i/2$ pairs $(k_i,l_i)$ into $\check K$. 
\epf

For example, for $K = \big((1,2),(1,3),(1,3),(1,3),(2,3),(2,3),(2,3),(2,3)\big)$, we have
$$
\hat K = \big((1,2),(1,3)\big)
 \,,\quad
\check K = \big((1,3),(2,3),(2,3)\big)\,.
$$
Define 
 \ala{
e_{(I,K,L)} 
 & :=
\prod_{i \in I} t_i
 \,
\prod_{(k_1,k_2) \in K} t_{k_1 k_2}
 \,
\prod_{(l_1,l_2,l_3) \in L} t_{l_1 l_2 l_3}
 \,,\quad
(I,K,L) \in \Sigma_1 \times \Sigma_2 \times \Sigma_3\,,
\\
p_{(K,L)} 
 & :=
\prod_{(k_1,k_2) \in \check K} p^T_{k_1 k_2}
 \,\,
\prod_{(l_1,l_2,l_3) \in L} p^T_{l_1 l_2 l_3}
 \,,\quad
(K,L) \in \Sigma_2 \times \Sigma_3\,,
 }
where by convention a product over an empty set yields $1$. By the first and the second fundamental theorem for invariants of complex matrices \cite[Thm.\ 3.4(a) and Cor.\ 4.4(a)]{Procesi}, as well as the relation \eqref{G-CHT} which follows from the Cayley-Hamilton theorem and holds true for matrices of determinant $1$, the set 
$$
B_0 := \left\{e_{(I,K,L)} : (I,K,L) \in \Sigma_1 \times \Sigma_2 \times \Sigma_3\right\}
$$
is a basis of the vector space $\mc R$. Denoting the length of a sequence by $|\cdot|$, we define
 \ala{
B
 & := 
\left\{e_{(I,\hat K,\varnothing)} p_{(\check K,L)}
 : 
(I,K,L) \in  \Sigma_1 \times \Sigma_2 \times \Sigma_3\right\}
\\
B_{\mc I}
 & := 
 \left\{
e_{(I,\hat K,\varnothing)} p_{(\check K,L)}
 : 
(I,K,L) \in  \Sigma_1 \times \Sigma_2 \times \Sigma_3 \,,\, |\check K| + |L| > 0
 \right\}
 }
and let $\mc X$ denote the span of $B_{\mc X} := B \setminus B_{\mc I}$.

\ble\label{L-Basis}~

\ben

\sitem\label{i-L-Basis-R} $B$ is a basis in $\mc R$.

\sitem\label{i-L-Basis-I} $B_{\mc I}$ is a basis in $\mc I$.

\sitem\label{i-L-Basis-X} $\mc X$ is a vector space complement of $\mc I$ in $\mc R$.

\een

\ele

\bpf

Point \rref{i-L-Basis-X} follows from points \rref{i-L-Basis-R} and \rref{i-L-Basis-I}.

\rref{i-L-Basis-R}.\abs Let $e_{(I,K,L)} \in B_0$ be given. For convenience, we will refer to the pair of nonnegative integers $(|K|,|L|)$ as the length of $e_{(I,K,L)}$. We can write 
$$
e_{(I,K,L)}
 =
\prod_{i \in I} t_i
 \,
\prod_{(k_1,k_2) \in \hat K} t_{k_1 k_2}
 \,
\prod_{(k'_1,k'_2) \in \check K} t_{k'_1 k'_2}^2
 \,
\prod_{(l_1,l_2,l_3) \in L} t_{l_1 l_2 l_3}
$$
and use formulae \eqref{G-pij} and \eqref{G-pijk} to replace all factors $t_{k_1' k_2'}^2$ by $p^T_{k_1' k_2'}$ and all factors $t_{l_1 l_2 l_3}$ by $p^T_{l_1 l_2 l_3}$. This yields
\beq\label{G-epee}
e_{(I,\hat K,\varnothing)} \, p_{(\check K,L)}
 =
2^{|\check K| + |L|} \, e_{(I,K,L)} + R_{(I,K,L)}\,,
\eeq
where $R_{(I,K,L)}$ is a linear combination of elements of $B_0$ having strictly smaller length than  $e_{(I,K,L)}$. Now, iterated application of this formula renders $e_{(I,K,L)}$ as a linear combination of the elements of $B$. This shows that $B$ spans $\mc R$. 

On the other hand, given a vanishing linear combination of the elements of $B$, we use \eqref{G-epee} to rewrite it as a linear combination of the elements of $B_0$. In the latter, the coefficients of the elements of largest length coincide up to multiplication by a power of $2$ with the coefficients of the corresponding elements of $B_0$. Hence, each of them must vanish, and we remain with a linear combination of elements of $B_0$ of smaller length. Iterated application of this argument then yields that all coefficients must vanish. Thus, $B$ is linearly independent and hence a basis of $\mc R$. 

\rref{i-L-Basis-I}.\abs Clearly, $\mc I$ is spanned by all products of $h \in \mc R$ with $p_{(K,L)}$ for some $(K,L) \in \Sigma_2 \times \Sigma_3$ such that $|K| + |L| > 0$. By point \rref{i-L-Basis-R}, we can expand $h$ with respect to the basis $B$. The assertion now follows by observing that for any $(I',K',L') \in \Sigma_1 \times \Sigma_2 \times \Sigma_3$, one has
$$
\big(e_{(I',\hat K',\varnothing)} \,\, p_{(\check K',L')} \big) p_{(K,L)}
 =
e_{(I',\hat K',\varnothing)} \, p_{(\check K' \sqcup K , L' \sqcup L)}
 =
e_{(I',\hat K'',\varnothing)} \, p_{(\check K'',L'')}
\,,
$$
where $K'' = K' \sqcup K \sqcup K$ and $L'' = L' \sqcup L$, which shows that the basis $B$ is invariant under multiplication by $p_{(L,K)}$.
\epf

As a result, $\mc X$ is spanned by the basis elements
$$
b_{(I,K)} := e_{(I,K,\varnothing)}
 \,,\qquad
(I,K) \in \Sigma_1 \times \widehat\Sigma_2\,,
$$ 
and we can expand
\beq\label{G-Entw}
f_{\mc X}
 =
\sum\nolimits_{(I,K) \in \Sigma_1 \times \widehat\Sigma_2} f_{(I,K)} \, b_{(I,K)}.
\eeq

\subsection{An approximate multiplication formula}

To analyze condition \eqref{G-Bed}, we need an approximate multiplication formula for the basis elements $b_{(I,K)}$. This will be derived now. 

We start with introducing some notation. For $K = \big((k_{11},k_{12}) , \dots , (k_{r1},k_{r2})\big) \in \widehat\Sigma_2$, let $\ol K \in \Sigma_1$ denote the sequence obtained from $(k_{11},k_{12} , \dots , k_{r1},k_{r2})$ by reordering. For $(I,K) \in \Sigma_1 \times\widehat\Sigma_2$ and $i = 1 , \dots , N$, let $\deg_i(I,K)$ count how many times the number $i$ appears in $I$ and in the pairs constituting $K$. We define the degree of $(I,K)$ by 
$$
\deg(I,K) := \big(\deg_1(I,K) , \dots , \deg_N(I,K)\big)\,.
$$
Moreover, we define the degree $\deg(h)$ of an element $h \in \mc R$ to be the maximum, with respect to the lexicographic ordering, of $\deg(I,K)$ over all elements $e_{(I,K,L)}$ of the basis $B_0$ appearing in the expansion of $h$ with respect to that basis with a nontrivial coefficient. We have
 \al{\label{G-deg-Pr}
\deg(h_1 h_2) & = \deg(h_1) + \deg(h_2)\,,
\\ \label{G-deg-Su}
\deg(h_1 + h_2) & \leq \max\left\{\deg(h_1) , \deg(h_2)\right\}\,.
 }
Finally, we observe that the elements of $\widehat\Sigma_2$ may be identified with subsets of the set of strongly increasing pairs of the numbers $1 , \dots , N$. Hence, given $K , K' \in \widehat\Sigma_2$, we may take the intersection $K \cap K'$ (the ordered sequence of pairs that $K$ and $K'$ have in common) and the union $K \cup K'$ (the ordered sequence of pairs appearing in $K$ or $K'$, where each pair that $K$ and $K'$ have in common appears just once). We have
\beq\label{G-XOR}
K \sqcup K' = (K \cup K') \sqcup (K \cap K')
\,,\quad
(K \ul\cup K') \sqcup (K \cap K') = K \cup K'\,,
\eeq
where $K \ul\cup K' := (K \setminus K') \cup (K' \setminus K)$ denotes the exclusive union (XOR). Using the operations of intersection and union, we can define an operation on $\Sigma_1 \times \widehat\Sigma_2$ by 
$$
(I,K) \cdot (I',K') := (I \sqcup I' \sqcup \ol{K \cap K'} , K \cup K')\,.
$$
As a consequence of the first formula in \eqref{G-XOR}, 
\beq\label{G-deg-K}
\deg\big((I,K) \cdot (I',K')\big)
 = 
\deg(I,K) + \deg(I',K')\,.
\eeq

\bbs

For 
$$
I = (1,3) \,,~~ K = \big((1,2),(1,4),(2,3)\big)
\,,~~
I' = (2) \,,~~ K' = \big((1,2),(1,3),(2,3)\big)\,, 
$$
we obtain $K \cap K' = \big((1,2),(2,3)\big)$ and thus $\ol{K \cap K'} = (1,2,2,3)$. Consequently, 
$$
I \sqcup I' \sqcup \ol{K \cap K'} = (1,1,2,2,2,3,3)
 \,,\quad
K \cup K' = \big((1,2),(1,3),(1,4),(2,3)\big)\,.
$$
By counting members, one may confirm \eqref{G-deg-K}.
\qeb

\ebs

\ble\label{L-Pr}

For all $(I,K) , (I',K') \in \Sigma_1 \times \widehat\Sigma_2$, we have the approximate multiplication formula
\beq\label{G-bb}
b_{(I,K)} \, b_{(I',K')} 
 =
b_{(I,K) \cdot (I',K')} + Q + R\,,
\eeq
where $Q \in \mc I$, $R \in \mc X$ and $\deg(R) < \deg\big(b_{(I,K) \cdot (I',K')}\big) = \deg(I,K) + \deg(I',K')$.

\ele

\bpf

We calculate
\beq\label{G-L-Pr-1}
b_{(I,K)} b_{(I',K')}
  =
\prod_{i \in I \sqcup I'} t_i
 \,
\prod_{(k_1,k_2) \in K \sqcup K'} t_{k_1 k_2}
 =
b_{(I \sqcup I' , K \ul\cup K')}
 \,
\prod_{(k_1,k_2) \in K \cap K'} t_{k_1 k_2}^2
\eeq
According to \eqref{G-pij},
$
t_{k_1 k_2}^2
 =
\frac 1 2 p^T_{k_1 k_2} + t_{k_1} t_{k_2} t_{k_1 k_2} - t_{k_1}^2 - t_{k_2}^2 + 4\,.
$
Hence,
$$
\prod\nolimits_{(k_1,k_2) \in K \cap K'} t_{k_1 k_2}^2
 =
\prod\nolimits_{(k_1,k_2) \in K \cap K'} t_{k_1} t_{k_2} t_{k_1 k_2}
 +
Q_1 + R_1\,,
$$
where $Q_1 \in \mc I$ and $R_1 \in \mc X$, and where 
\beq\label{G-L-Pr-2}
\deg(R_1) < 2 \deg(\varnothing , K \cap K')\,.
\eeq
Plugging this into \eqref{G-L-Pr-1} and using the second formula in \eqref{G-XOR}, we obtain
$$
b_{(I,K)} \, b_{(I',K')}
 =
b_{(I,K) \cdot (I',K')}
 +
Q_2
 + 
b_{(I \sqcup I' , K \ul\cup K')} \, R_1
 \,,
$$
where $Q_2 \in \mc I$. Writing $b_{(I \sqcup I' , K \ul\cup K')} \, R_1 = Q_3 + R$ with $Q_3 \in \mc I$ and $R \in \mc X$, we arrive at \eqref{G-bb} with $Q = Q_2 + Q_3$. It remains to compare the degrees. By \eqref{G-deg-Su}, we have $\deg(R) \leq \deg\big(b_{(I \sqcup I' , K \ul\cup K')} \, R_1\big)$. Moreover, \eqref{G-deg-Pr} and \eqref{G-L-Pr-2} imply
$$
\deg\big(b_{(I \sqcup I' , K \ul\cup K')} \, R_1\big)
 <
\deg(I \sqcup I' , K \ul\cup K') + 2 \deg(\varnothing , K \cap K')\,.
$$
It is easy to see that the right hand side equals $\deg(I,K) + \deg(I',K')$.
\epf

Now, we use the approximate multiplication formula \eqref{G-bb} for showing that condition \eqref{G-Bed} implies $f_{\mc X} = 0$. For that purpose, recall the expansion of $f_{\mc X}$ given by \eqref{G-Entw}. In what follows, let $\NN_0$ denote the set of nonnegative integers. For $\mu \in \NN_0^N$, let 
$$
\Sigma_\mu := \{(I,K) \in \Sigma_1 \times \widehat\Sigma_2 : \deg(I,K) = \mu\}\,.
$$

\bbs\label{B-Sm}

For $N=3$ and $\mu = (2,1,1)$, $\Sigma_\mu$ consists of the following elements:
 \ala{
 &&
(I_0,K_0) & = \big((1,1,2,3),\varnothing\big)
 \,, &
(I_1,K_1) & = \big((1,2),\big((1,3)\big)\big)
 \,,
&&
\\
&&
(I_2,K_2) & = \big((1,3),\big((1,2)\big)\big)
 \,, &
(I_3,K_3) & = \big((1,1),\big((2,3)\big)\big)\,,
&&
\\
&& (I_4,K_4) & = \big(\varnothing,\big((1,2),(1,3)\big)\big)\,. &&
 }
\qeb
\ebs

Given $\mu \in \NN_0^N$ and $(I,K) \in \Sigma_{2\mu}$, let 
$$
F_{(I,K)}
 :=
\sum_{(I',K') , (I'',K'') \in \Sigma_\mu \atop (I',K') \cdot (I'',K'') = (I,K)} f_{(I',K')} \, f_{(I'',K'')}\,,
$$
For every $\mu \in \NN_0^N$, consider the following two statements.
\medskip

\abs$(\mr A_\mu)$\abs
{\it $F_{(I,K)} = 0$ for all $(I,K) \in \Sigma_{2\mu}$.}
\medskip

\abs$(\mr B_\mu)$\abs
{\it $f_{(I,K)} = 0$ for all $(I,K) \in \Sigma_\mu$.}

\ble\label{L-Itr}

If $(\mr A_\mu)$ implies $(\mr B_\mu)$ for all $\mu \in \N_0^N$, then $f_{\mc X} = 0$. 

\ele

\bpf

In the first step, we choose $\mu \in \NN_0^N$ such that $\deg(f_{\mc X}) \leq \mu$. By \eqref{G-deg-Pr} and \eqref{G-deg-Su}, then $\deg\big(\big(f_{\mc X}^2\big)_{\mc X}\big) \leq 2\mu$. As a consequence, Lemma \rref{L-Pr} and formula \eqref{G-deg-K} yield that the contribution to $\big(f_{\mc X}^2\big)_{\mc X}$ carried by the elements of $B_{\mc X}$ of degree $2\mu$ is given by
$$
\sum_{(I,K) , (I',K') \in \Sigma_\mu} f_{(I,K)} \,\, f_{(I',K')} \,\, b_{(I,K) \cdot (I',K')}\,.
$$
According to \eqref{G-Bed}, this sum vanishes. Sorting by basis elements, we find that $F_{(I'',K'')} = 0$ for all $(I'',K'') \in \Sigma_{2\mu}$. Thus, condition $(\mr A_\mu)$ holds true for the $N$-tuple $\mu$ under consideration. By assumption, then $(\mr B_\mu)$ holds true. It follows that $\deg(f_{\mc X}) < \mu$. Now, the argument can be iterated by decrementing $\mu$. As a result, we find that $f_{(I,K)} = 0$ holds for all $(I,K) \in \Sigma_1 \times \widehat\Sigma_2$. Hence, $f_{\mc X} = 0$.
\epf

\subsection{Condition $(\mr A_\mu)$ implies $(\mr B_\mu)$}

Let $\mu \in \NN_0^N$ be chosen. Given a subset $\mc K \subset \Sigma_\mu$, we will write 
$$
S(\mc K) := \sum_{(I,K) \in \mc K} f_{(I,K)}\,.
$$
First, we observe that 
$$
S(\Sigma_\mu)^2
 =
\sum_{(I,K) , (I',K') \in \Sigma_\mu} f_{(I,K)} f_{(I',K')}
 =
\sum_{(I'',K'') \in \Sigma_{2\mu}} F_{(I'',K'')}\,.
$$
By condition $(\mr A_\mu)$, this vanishes. Hence,
\beq\label{G-Smu}
S(\Sigma_\mu) = 0\,.
\eeq
Now, for every $J \in \widehat\Sigma_2$, we define
$$
\mc K_J^i := \{(I,K) \in \Sigma_\mu : |K \cap J| = i\}\,.
$$

\bbs

We take up Example \rref{B-Sm}. For $J = \big((1,2),(1,3)\big)$, we obtain
$$
\mc K_J^0 = \{(I_0,K_0),(I_3,K_3)\}
 \,,\quad
\mc K_J^1 = \{(I_1,K_1),(I_2,K_2)\}
 \,,\quad
\mc K_J^2 = \{(I_4,K_4)\}
$$
and $K_J^i = \varnothing$ for $i > 2$.
\qeb

\ebs

\ble\label{L-KLi}

For every $J \in \widehat\Sigma_2$ and every $i \in \NN_0$, one has $S(\mc K^i_J) = 0$.

\ele

\bpf

We prove the assertion by induction on the length of $J$. Since it is obvious for $i > |J|$, we may assume $i \leq |J|$ throughout. Still, depending on $\mu$, it might happen that $\mc K^i_J = \varnothing$ for some $J$. This has no effect on the argument.

The base case is $|J| = 0$, that is, $J = \varnothing$. Here, $i=0$ and hence  $\mc K^i_J = \Sigma_\mu$, so that the assertion follows from \eqref{G-Smu}.

To accomplish the inductive step, let $J \in \widehat\Sigma_2$ be given, let $r = |J|$ and assume that the assertion holds for all $J' \in \widehat\Sigma_2$ of length $|J'| < r$. First, we will show that $S(\mc K^i_J)$ is proportional to $S(\mc K^r_J)$. Write $J = (P_1 , \dots , P_r)$ with pairs $P_j$. For $1 \leq j \leq r$, let $J_j$ denote the sequence obtained from $J$ by omitting $P_j$. Consider the sum
\beq\label{G-L-KLi-1}
\sum_{j = 1}^r S\big(\mc K_{J_j}^i \setminus \mc K^1_{(P_j)}\big)\,.
\eeq
On the one hand, if for a given $(I,K) \in \Sigma_\mu$ one has $K \cap J = (P_{j_1} , \dots , P_{j_i})$, then the coefficient $f_{(I,K)}$ appears in the summands where $j \neq j_1 , \dots , j_i$. Hence, the sum \eqref{G-L-KLi-1} equals $(r-i) S\big(\mc K^i_J\big)$. On the other hand, this sum can be rewritten as 
$$
\sum_{j = 1}^r
 \Big[
S\big(\mc K^i_{J_j}\big) - S\big(\mc K^i_{J_j} \cap \mc K^1_{(P_j)}\big)
 \Big]\,.
$$
In each summand, the first term vanishes by the induction assumption. It follows that
\beq\label{G-L-KLi-2}
S\big(\mc K^i_J\big)
 =
- \frac{1}{r-i} \sum_{j = 1}^r
S\big(\mc K^i_{J_j} \cap \mc K^1_{(P_j)}\big)\,.
\eeq
Now, for each $j$, 
\beq\label{G-L-KLi-4}
S\big(\mc K^i_{J_j} \cap \mc K^1_{(P_j)}\big)
 =
S\big(\mc K^{i+1}_J \setminus \mc K^{i+1}_{J_j}\big)
 =
S\big(\mc K^{i+1}_J\big) - S\big(\mc K^{i+1}_J \cap \mc K^{i+1}_{J_j}\big)\,.
\eeq
Using $\mc K^{i+1}_J \cap \mc K^{i+1}_{J_j} = \mc K^{i+1}_{J_j} \setminus \mc K^1_{(P_j)}$, the second term can be rewritten as 
\beq\label{G-L-KLi-5}
S\big(\mc K^{i+1}_J \cap \mc K^{i+1}_{J_j}\big)
 =
S\big(\mc K^{i+1}_{J_j}\big) - S\big(\mc K^{i+1}_{J_j} \cap \mc K^1_{(P_j)}\big)
 =
- S\big(\mc K^{i+1}_{J_j} \cap \mc K^1_{(P_j)}\big)\,,
\eeq
where the second equality is due to the induction assumption. Plugging \eqref{G-L-KLi-5} into \eqref{G-L-KLi-4}, we obtain
$$
S\big(\mc K^i_{J_j} \cap \mc K^1_{(P_j)}\big)
 =
S\big(\mc K^{i+1}_J\big) + S\big(\mc K^{i+1}_{J_j} \cap \mc K^1_{(P_j)}\big)\,.
$$
Iterating this, we obtain
\beq\label{G-L-KLi-3}
S\big(\mc K^i_{J_j} \cap \mc K^1_{(P_j)}\big)
 =
S\big(\mc K^{i+1}_J\big) + \cdots + S\big(\mc K^r_J\big)\,,
\eeq
because $\mc K^r_{J_j} = \varnothing$. Consequently, \eqref{G-L-KLi-2} yields 
$$
S\big(\mc K^i_J\big)
 =
- \frac{r}{r-i} \left[S\big(\mc K^{i+1}_J\big) + \cdots + S\big(\mc K^r_J\big)\right]\,,
$$
from which we iteratively conclude that $S\big(\mc K^i_J\big)$ is proportional to $S\big(\mc K^r_J\big)$, indeed. More precisely, we obtain
$$
\textstyle
S\big(\mc K^i_J\big)
 =
(-1)^{r-i} \, {r \choose i} \, S\big(\mc K^r_J\big)\,.
$$
Now, we use this and the disjoint decomposition 
$$
\Sigma_\mu
 = 
\mc K^0_J \cup \mc K^1_J \cup \cdots \cup \mc K^r_J
$$
to see that $S(\mc K^r_J)$ is proportional to $S(\Sigma_\mu)$ and hence, by \eqref{G-Smu}, that $S\big(\mc K^r_J\big) = 0$.
\epf

To complete the argument that condition $(\mr A_\mu)$ implies condition $(\mr B_\mu)$, we observe that for each $(I,K) \in \Sigma_\mu$, the sequence $I$ is uniquely determined by $K$. Let 
$$
m := \max{}\big\{|K| : (I,K) \in \Sigma_\mu\big\}\,.
$$
For every $(I,K) \in \Sigma_\mu$ with $|K| = m$, one has $\mc K^m_K = \{(I',K') \in \Sigma_\mu : K' = K\}$. Since $I'$ is determined by $K'$ and the degree $\mu$, we conclude that $\mc K^m_K = \{(I,K)\}$. Hence, $S(\mc K^m_K) = f_{(I,K)}$ and Lemma \rref{L-KLi} yields $f_{(I,K)} = 0$. In turn, for $(I,K) \in \Sigma_\mu$ with $|K| = m-1$, we have $\mc K^{m-1}_K = \{(I',K') \in \Sigma_\mu : K \subset K'\}$. By the same argument as above we conclude that $\mc K^{m-1}_K$ consists of $(I,K)$ itself and elements $(I',K') \in \Sigma_\mu$ with $|K'| > |K|$. Hence, 
$$
S\big(\mc K^{m-1}_K\big)
 = 
f_{(I,K)} + \sum_{(I',K') \in \Sigma_\mu \atop |K'| = m} f_{(I',K')}
 =
f_{(I,K)}
$$
and thus $f_{(I,K)} = 0$. Iterating this argument, we finally obtain $f_{(I,K)} = 0$ for all $(I,K) \in \Sigma_\mu$. It follows that $(\mr A_\mu)$ implies $(\mr B_\mu)$. In view of Lemma \rref{L-Itr}, this completes the proof of Theorem \rref{T-RI}.
\qed

\section{Summary and outlook}

To summarize, in order to find the vanishing subspace $\mc V_\tau$ associated with a stratum $\tau$ of the classical phase space of a gauge field model 
on a finite lattice with a compact gauge group $G$ one has to find a set of polynomial invariants satisfying two conditions: 
\begin{enumerate}
\item
the zero locus condition,
\item 
the radical ideal condition, 
\end{enumerate}
see Proposition \ref{S-V}. In this paper, we have constructed such a set for the torus stratum  $\pha_T$  of the $G = \SU(2)$-model. The remaining secondary strata 
consist of isolated points and thus the corresponding vanishing subspaces can be obtained in a straightforward way as in \cite{HRS}.  

In conclusion,  it remains to construct the subspaces $\mc V_T$ and $\mc H_T$ associated with
the torus stratum $\pha_T$ explicitly. According to Theorems \rref{T-Rel-T}, \rref{T-RI} and Corollary \rref{F-V},
$$
\mc V_T
 ~=~
\left(\,\underset{i < j}{\text{\LARGE$\mathbf +$}}
\im\big(\hat p^T_{ij}\big)\right)
 +
\left(\,\underset{i < j < k}{\text{\LARGE$\mathbf +$}} \,
\im\big(\hat p^T_{ijk}\big)\right)
$$
and 
$$
\mc H_T
 =
\left(\,\,\bigcap_{i < j} \ker\big((\hat p^T_{ij})^\dagger\big)\right)
 \cap
\left(\,\,\bigcap_{i < j < k} \ker\big((\hat p^T_{ijk})^\dagger\big)\right)\,,
$$
where $\hat p^T_{ij}$ and $\hat p^T_{ijk}$ denote the operators of
multiplication by the corresponding functions. Thus, one has to take an
orthonormal basis in $\mc H$ and to compute the matrix elements of these
operators. An orthonormal basis is given, for example, by the representative
functions 
$$
\chi_{l_1 , \dots , l_N ; l}(\ul a) 
 =
C_{l_1 , \dots , l_N ; l}
\tr
 \left(
P_{l_1 , \dots , l_N}^l
\big(\pi_{l_1}(a_1) \otimes \cdots \otimes \pi_{l_N}(a_N)\big)
 \right)\,,
$$
where $l_1 , \dots , l_N$ and $l$ are integers, $C_{l_1 , \dots , l_N ; l}$ is a normalization
constant, $\pi_{l_i}$ 
denotes the irreducible representation of $\SL(2,\CC)$ of spin $l_i/2$ and
$P_{l_1 , \dots , l_N}^l$ denotes the projector to the subrepresentation of
spin $l/2$ of the tensor product representation $\pi_{l_1} \otimes \cdots
\otimes \pi_{l_N}$. Thus, for given $l_1 , \dots , l_N$, the range of $l$ is
restricted by the condition that $\pi_l$ occurs as a subrepresentation of
$\pi_{l_1} \otimes \cdots \otimes \pi_{l_N}$. To find the matrix elements of
$\hat p^T_{ij}$ and 
$\hat p^T_{ijk}$, one has to expand, respectively, $p^T_{ij} \chi_{s_1 , \dots , s_N ; s}$ and
$p^T_{ijk} \chi_{s_1 , \dots , s_N ; s}$ in that basis. This is a
problem in the representation theory of $\SL(2,\CC)$, which will be addressed in the future.

\section*{Acknowledgements}

We are greatly indebted to the referee for outlining the arguments in Section \rref{A-SQT-AA-defRel} and for encouraging us to clarify whether the ideal generated by the relations defining the secondary stratum $\pha_T$ is a radical ideal (Theorem \rref{T-RI}).


\begin{thebibliography}{99}

\bibitem{AbrahamMarsden}
 R.\ Abraham, J.E.\ Marsden: 
 {\em Foundations of Mechanics.}
 Benjamin/Cummings 1978

\bibitem{cfg}
 S.\ Charzy\'nski, J.\ Kijowski, G.\ Rudolph, M.\ Schmidt:
 On the stratified classical configuration space of lattice QCD.
 J.\ Geom.\ Phys.\ {\bf 55} (2005) 137--178

\bibitem{cfgtop}
 S.\ Charzy\'nski, G.\ Rudolph, M.\ Schmidt:
 On the topological structure of the stratified classical configuration space of lattice QCD. 
 J.\ Geom.\ Phys. {\bf 58} (2008) 1607--1623

\bibitem{FRS}
 E.\ Fischer, G.\ Rudolph, M.\ Schmidt:
 A lattice gauge model of singular Marsden-Weinstein reduction. Part I. Kinematics. 
 J.\ Geom.\ Phys.\ {\bf 57} (2007) 1193--1213

\bibitem{GR}
 H.\ Grundling, G.\ Rudolph:  
 QCD on an infinite lattice. 
 Commun.\ Math.\ Phys.\ {\bf 318} (2013)  717--766
 
\bibitem{GR2}
 H.\ Grundling, G.\ Rudolph:  
 Dynamics for QCD on an infinite lattice. 
 Commun.\ Math.\ Phys.\ (2016) doi:10.1007/s00220-016-2733-5
 
\bibitem{Hall:SBT}
 B.C.\ Hall: 
 The Segal-Bargmann ''coherent state'' transform for compact Lie groups.
 J.\ Funct.\ Anal.\ {\bf 122} (1994) 103--151

\bibitem{Hall:cptype}
 B.C.\ Hall: 
 Geometric quantization and the generalized Segal-Bargmann transform
 for Lie groups of compact type.
 Commun.\ Math.\ Phys.\ \textbf{226} (2002) 233--268

\bibitem{HeinznerLoose}
 P.\ Heinzner, F.\ Loose:
 Reduction of complex Hamiltonian $G$-spaces.
 Geom.\ Funct.\ Anal.\ {\bf 4} (1994) 288--297

\bibitem{adg}
 M.\ Hofmann, G.\ Rudolph, M.\ Schmidt:
 Orbit type stratification of the adjoint quotient of a compact semisimple Lie group.
 J.\ Math.\ Phys.\ {\bf 54} (2013) 083505

\bibitem{Hue:Quantization}
 J.\ Huebschmann:
 K\"ahler quantization and reduction.
 J.\ reine angew.\ Math.\ \textbf{591} (2006) 75--109,
 {\tt math.SG/0207166}

\bibitem{HRS}
 J.\ Huebschmann, G.\ Rudolph, M.\ Schmidt: 
 A lattice gauge model for quantum mechanics on a stratified space.
 Commun.\ Math.\ Phys. {\bf 286} (2009) 459--494

\bibitem{qcd1}
 P.D.\ Jarvis, J.\ Kijowski, G.\ Rudolph:
 On the structure of the observable algebra of QCD on the lattice.
 J.\ Phys.\ A {\bf 38} (2005) 5359--5377
 
\bibitem{KempfNess}
 G.\ Kempf, L.\ Ness: 
 The length of vectors in representation spaces.
 In: {\it Algebraic Geometry}, Lect.\ Notes Math.\ {\bf 732}, Springer 1979, pp.\ 233--244
 
\bibitem{qcd2}
 J.\ Kijowski, G.\ Rudolph:
 On the Gauss law and global charge for quantum chromodynamics.
 J.\ Math.\ Phys.\ {\bf 43} (2002) 1796--1808

\bibitem{qcd3}
 J.\ Kijowski, G.\ Rudolph:
 Charge superselection sectors for qcd on the lattice.
 J.\ Math.\ Phys.\ {\bf 46} (2005) 032303

\bibitem{Mumford}
 D.\ Mumford, J.\ Fogarty, F.\ Kirwan:
 {\em Geometric Invariant Theory.}
 Springer 1994
 
\bibitem{OrtegaRatiu}
 J.-P.\ Ortega, T.S.\ Ratiu: 
 {\em Momentum Maps and Hamiltonian Reduction.}
 Progress in Mathematics, Vol.\ 222, Birkhäuser 2004

\bibitem{Procesi}
 C.\ Procesi:
 The invariant theory of $n\times n$ matrices.
 Adv.\ Math.\ {\bf 19} (1976) 306--381

\bibitem{Procesi:LG}
 C.\ Procesi:
 {\em Lie Groups.}
 Universitext, Springer 2007

\bibitem{RS}
 G.\ Rudolph, M.\ Schmidt: 
 On the algebra of quantum observables for a certain gauge model. 
 J.\ Math.\ Phys. {\bf 50} (2009) 052102
 
\bibitem{Buch}
 G.\ Rudolph, M.\ Schmidt: 
 {\it Differential Geometry and Mathematical Physics. Part I. Manifolds, Lie Groups and Hamiltonian Systems.}
 Springer 2013

\bibitem{Serre}
 J.-P.\ Serre:
 G\'eom\'etrie alg\'ebrique et g\'eom\'etrie analytique.
 Ann.\ Inst.\ Fourier {\bf 6} (1956) 1--42

\bibitem{SjamaarLerman}
 R.\ Sjamaar, E.\ Lerman:
 Stratified symplectic spaces and reduction.
 Ann.\ of Math.\ {\bf 134} (1991) 375--422

\end{thebibliography}
\end{document}